\documentclass[11pt]{article}
\usepackage[utf8]{inputenc}
\usepackage{amsmath}
\usepackage{graphicx}
\usepackage{caption}
\usepackage{amssymb}
\usepackage[hidelinks]{hyperref}
\usepackage{geometry}
\usepackage{textcomp}
\usepackage[numbers,square]{natbib}

\title{
\Large Resonance Complexity Theory and the Architecture of Consciousness:\\
\normalsize A Field-Theoretic Model of Resonant Interference and Emergent Awareness
}
\author{Michael Arnold Bruna}
\date{May 17, 2025}

\usepackage{float}       
\usepackage[section]{placeins} 

\begin{document}

\maketitle

\begin{center}
\textbf{Affiliation:} Independent Researcher \

\textbf{Date:} May 17, 2025 \

\textbf{Contact:} \href{mailto:Michael.Bruna88@gmail.com}{Michael.Bruna88@gmail.com}

\end{center}

\begin{abstract}
This paper introduces Resonance Complexity Theory (RCT), which proposes that consciousness emerges from stable interference patterns of oscillatory neural activity. These patterns, shaped by recursive feedback and constructive interference, must exceed critical thresholds in complexity, coherence, gain, and fractal dimensionality to give rise to conscious experience. The resulting spatiotemporal attractors encode subjective awareness as dynamic resonance structures distributed across the neural field, enabling large-scale integration without symbolic representation or centralized control.

To formalize this idea, we define the Complexity Index (CI), a composite metric that synthesizes four core properties of conscious systems: fractal dimensionality ($D$), signal gain ($G$), spatial coherence ($C$), and attractor dwell time ($\tau$). These elements are combined multiplicatively to capture the emergence and persistence of structured, integrative neural states.

To test the theory empirically, we developed a biologically inspired yet minimal neural field simulation composed of radial wave sources emitting across a continuous 2D space. The system exhibits recursive constructive interference, producing coherent, attractor-like excitation patterns without external input, regional coding, or imposed structure. These patterns meet the theoretical thresholds for CI and reflect the core dynamics predicted by RCT.

The findings demonstrate that resonance-based attractors—and by extension, consciousness-like dynamics—can arise purely from the physics of wave interference. RCT thus offers a unified, dynamical framework for modeling awareness as an emergent property of organized complexity in oscillatory systems.
\end{abstract}

\section{The Resonance Complexity Theory}

Consciousness. The Hard Problem. For centuries, philosophers, physicists, and neuroscientists have sought to explain how subjective experience emerges from physical processes, yet a complete account remains elusive. Resonance Complexity Theory (RCT) proposes that consciousness arises from the self-organization of oscillatory neural activity into stable, resonant patterns. These patterns—manifesting as spatiotemporal attractors—emerge through constructive interference and recurrent feedback, and must exceed critical thresholds of complexity, coherence, gain, and persistence to support awareness.

To quantify this phenomenon, we define the \textit{Complexity Index (CI)}:
\begin{equation}
\text{CI} = \alpha D \cdot G \cdot C \cdot \left(1 - e^{-\beta \tau} \right)
\end{equation}
where $D$ is fractal dimensionality, $G$ is signal gain, $C$ is spatial coherence, $\tau$ is attractor dwell time, and $\alpha$, $\beta$ are scaling constants. All components must co-occur to produce conscious-level resonance: complex, coherent, and temporally stable.

Unlike symbolic or representational accounts of consciousness, RCT offers a physical mechanism grounded in field dynamics. It emphasizes the interplay of recursive resonance, phase synchrony, fractal structure, and feedback-driven amplification to stabilize meaningful global patterns from distributed activity. These stabilized waveforms form a unified experiential substrate—a dynamic present state we interpret as consciousness.

Although not dependent on thermodynamic entropy, RCT predicts that conscious states will exhibit high informational richness. Shannon entropy \citep{Shannon1948} serves as a useful complementary measure, distinguishing repetitive or low-dimensional activity from richly differentiated attractor structures.

To evaluate RCT, we developed a minimal neural field simulation consisting of distributed radial wave sources across a 2D medium. Without any external input or imposed architecture, the system exhibits spontaneous attractor formation through recursive constructive interference. The resulting excitation patterns align with RCT’s theoretical thresholds for CI, demonstrating that consciousness-like dynamics can emerge from the physics of wave interaction alone.

These findings support RCT’s core hypothesis: that awareness arises not from symbolic encoding or hierarchical access, but from resonance itself—stabilized, structured, and self-sustaining across time.

\section{Background and Related Work}

Numerous theoretical frameworks have shaped the scientific study of consciousness, each offering a unique lens on how awareness arises. These can be broadly grouped into informational, access-based, predictive, and dynamical models.

\textit{Integrated Information Theory (IIT)} \citep{Tononi2004,Tononi2016} defines consciousness as the capacity of a system to integrate information beyond the sum of its parts, quantified by the scalar $\Phi$. IIT offers a formal structure and testable predictions, but remains agnostic about how high-$\Phi$ states emerge dynamically in neural systems. Its recent iterations add causal structure and spiking models \citep{Tononi2022}, but still lack a mechanistic substrate for conscious integration.

\textit{Global Workspace Theory (GWT)} \citep{Baars1988,Chalmers1995} frames consciousness as information broadcast across a modular network. \textit{The Free Energy Principle (FEP)} \citep{Friston2010} provides a predictive coding framework in which brains minimize surprise. Both models are powerful in scope but do not directly address the dynamical formation of stable, structured states that underlie experience.

Other accounts, including \textit{Recurrent Processing Theory (RPT)} and \textit{Higher-Order Theories (HOT)}, emphasize feedback and meta-representation, but similarly stop short of modeling the underlying physics of neural field dynamics.

RCT complements these approaches by proposing a concrete biophysical mechanism: that consciousness emerges from constructive interference among nested oscillations across multiple frequencies. These interactions form spatiotemporal attractors stabilized by recursive feedback and coherence. RCT offers a field-based account that explains not just access or integration, but the emergence of structured awareness itself.

Rather than replacing existing frameworks, RCT provides a dynamical foundation on which informational and representational theories may rest, offering a mechanistic pathway through which integration, access, and awareness can be realized in the brain's physical substrate.

\section{Core Theory}

This work proposes that consciousness emerges from the formation of stable, self-sustaining attractors within the brain's oscillatory field dynamics. These attractors arise through constructive interference among neural oscillations distributed across multiple spatial and temporal scales.

In classical dynamical systems theory, an attractor is a region in phase space toward which a system tends to evolve over time. RCT generalizes this concept to describe structured spatiotemporal interference patterns—dynamic resonance formations that persist through recursive wave interaction. These field-based attractors are not fixed points but coherent, self-reinforcing geometries that encode the instantaneous configuration of awareness.

When constructive interference exceeds critical thresholds of structure and persistence—quantified by measures of complexity, coherence, gain, and dwell time—subjective awareness emerges as a natural consequence of the system’s resonant organization.

Within this framework, attractors are dynamic spatiotemporal structures, shaped by phase alignment, recursive feedback, and cross-frequency coupling. Unlike models focused on symbolic transmission \citep{Shannon1948}, this work posits that the emergence and depth of conscious experience depend on the stability and recurrence of these harmonic formations.

To quantify this, we define the \textit{Complexity Index (CI)}, a unified metric that integrates four key variables:

\begin{equation}
\text{CI} = \alpha \cdot D \cdot G \cdot C \cdot \left(1 - e^{-\beta \cdot \tau}\right)
\end{equation}

Where:
\begin{itemize}
  \item \textbf{$D$ — Fractal Dimensionality:} Captures the nested spatial complexity of the resonance pattern. High $D$ indicates that the attractor spans multiple spatial scales and encodes information in a richly layered, self-similar structure.

  \item \textbf{$G$ — Signal Gain:} Reflects the overall energy or amplitude of oscillatory activity in the region. It measures how strongly a local area contributes to the wave dynamics, with higher $G$ indicating greater excitability or activation.

  \item \textbf{$C$ — Spatial Coherence:} Measures the degree of local synchrony across the field. High $C$ means that neighboring oscillatory units are aligned in phase and frequency, enabling constructive interference and the formation of unified resonance patterns.

  \item \textbf{$\tau$ — Dwell Time:} Represents the temporal stability of the attractor. It quantifies how long a particular resonant configuration persists over time, reflecting the system’s ability to maintain an internally consistent state.

  \item \textbf{$\alpha$, $\beta$ — Scaling Constants:} Normalization terms that control the overall range and sensitivity of CI. $\alpha$ sets the global scale of the index, while $\beta$ governs the contribution of $\tau$ through an exponential saturation curve.
\end{itemize}

The multiplicative form ensures that all components must co-occur to generate a conscious state; no single variable is sufficient on its own and CI collapses if any of these factors is absent.

\subsection{Recursive Formulation of CI}

Consciousness is inherently hierarchical. To capture this, we introduce a recursive formulation of CI, where each frequency band contributes meaningfully only if lower-frequency patterns are already stable:

\begin{equation}
CI^{(n)} = \alpha_n \cdot D^{(n)} \cdot CI^{(n-1)} \cdot C^{(n)} \cdot \left(1 - e^{-\beta_n \cdot \tau^{(n)}}\right)
\end{equation}

The base layer, typically corresponding to delta-band oscillations, is defined as:

\begin{equation}
CI^{(0)} = \alpha_0 \cdot D^{(0)} \cdot G^{(0)} \cdot C^{(0)} \cdot \left(1 - e^{-\beta_0 \cdot \tau^{(0)}}\right)
\end{equation}

This hierarchy reflects the principle that slower rhythms provide a scaffolding for higher-frequency dynamics, aligning with empirical evidence from cross-frequency coupling studies \citep{Canolty2006}.

\subsection{Summation Form and Unified CI}

While the recursive model emphasizes dependency, consciousness also draws from multiple bands in parallel. To accommodate this, we define a summation-based CI:

\begin{equation}
CI_{\text{total}} = \sum_{n=1}^{N} w_n \cdot \Phi_n
\end{equation}
\begin{equation}
\Phi_n = \alpha_n \cdot D_n \cdot \mathcal{C}_n \cdot \left(1 - e^{-\beta_n \cdot \tau_n} \right), \quad \mathcal{C}_n = C_n \cdot \prod_{k=0}^{n-1} CI_k
\end{equation}

To unify both recursive and additive contributions, we introduce the \textit{Unified CI} equation:

\begin{equation}
CI_{\text{total}} = \sum_{n=1}^{N} w_n \cdot \left[ \alpha_n \cdot D_n \cdot \left( \prod_{k=0}^{n-1} CI_k \right) \cdot C_n \cdot \left(1 - e^{-\beta_n \cdot \tau_n} \right) \right]
\end{equation}

This formalism models the brain as a \textit{nested interference lattice}—a hierarchy of resonant structures formed by constructive overlap across frequency bands. Each layer contributes to consciousness only when coherently embedded within slower, temporally stable scaffolds. This supports findings that high-frequency gamma activity is meaningful only when nested within lower-frequency rhythms \citep{Canolty2006, Buzsaki2006}.

\subsection{Theoretical Foundations of Resonance Complexity}

RCT states that consciousness arises naturally from the intrinsic tendency of the brain to form structured resonance patterns. These patterns emerge through constructive interference of neural oscillations across spatially distributed and hierarchically nested frequency bands. Unlike theories that rely on abstract measures of information integration or global broadcasting, our theory grounds consciousness in biophysically plausible mechanisms: oscillatory coupling, phase alignment, and recurrent feedback.

As oscillatory components interact, they give rise to stable self-organizing attractors that span anatomically and functionally distinct regions. These attractors are not imposed top-down but emerge spontaneously from local dynamics that cohere over time through recursive synchronization. The result is a multiscale lattice of resonant activity, where consciousness is defined by the degree to which such attractors are spatially structured, temporally stable, and recursively reinforced.

This foundation distinguishes RCT from purely informational models such as ITT \citep{Tononi2004}, and from cybernetic frameworks such as the free energy principle \citep{Friston2010}, by explicitly identifying the physical oscillatory mechanisms through which conscious structure and continuity emerge. In RCT, we do not deny the relevance of information integration or prediction; it subsumes them within a broader substrate of resonance-driven attractor dynamics that underpins the phenomenology of conscious awareness.

\subsection{Foundational Axioms of Resonance Complexity Theory}

To clarify the theoretical structure of this framework, we introduce a set of foundational axioms. These serve as the core assumptions from which the theory’s dynamics, metrics, and explanatory power are derived. The purpose of these axioms is twofold: (1) to explicitly state the ontological commitments of RCT, and (2) to provide a conceptual bridge between the formal mathematical framework (e.g., CI) and the phenomenological claims regarding subjective awareness.

\begin{enumerate}
    \item \textbf{Axiom 1: Interference generates stabilized wave patterns.} \\
    In an oscillatory medium, wavefronts propagate and interact through constructive and destructive interference. Under the right conditions, these interactions give rise to stable, self-sustaining interference patterns—structures that persist through recursive reinforcement. \citep{Chladni1961}
    This interference substrate forms the dynamic basis for higher-order organization.

    \item \textbf{Axiom 2: Attractors form from stabilized, recurrent wave patterns.} \\
    When interference patterns become recurrently stabilized through recursive feedback, they form spatiotemporal geometric biases—structures we call "attractors." These patterns bind distributed oscillatory neural activity into a unified, temporally persistent configuration. Attractors serve as the physical substrate upon which conscious states may form. Not all attractors are necessarily experienced. When sufficiently stable and coherent, the feedback loop provides the system a \textit{point of view} on its own activity.

    \item \textbf{Axiom 3: Attractors are layered across time and scale.} \\
    Attractors are hierarchically embedded across spatial and temporal scales. Higher-frequency oscillations contribute to awareness only when harmonically coupled to slower rhythms, enabling phase-locking and nested resonance. This layered structure binds fast, information-rich waveforms into slower, stabilizing scaffolds that support continuity of experience.

    \item \textbf{Axiom 4: Information is stored in wave geometry.} \\
    Conscious content is not encoded through spikes or discrete variability, but through the geometry of harmonic resonance. Meaning arises from the spatial patterning of interference, phase coherence, and recursive reinforcement—not from unstructured activity. Because this structure is intrinsically ordered and distributed across the field, it acts as a form of holographic compression—storing content not in localized codes, but in the geometry of resonance itself. The geometry of the attractor defines what is being experienced, and how compactly that experience can be maintained or recalled.

    \item \textbf{Axiom 5: The wave pattern is experience.} \\
    The attractor is not a representation of experience—it is experience itself. When a waveform becomes temporally stable, spatially coherent, and recursively integrated, it constitutes the system’s internal reality. The system does not observe the attractor—it is the attractor. It's recursive resonance, stability, and structure are what it feels like to be the system in that moment.
\end{enumerate}

\section{Mathematical Foundations of RCT}

RCT is grounded in a continuous field-theoretic model, where neural activity is represented as a superposition of oscillatory components propagating across a structured medium. This section introduces the formal mathematical underpinnings of the theory, including the definition of CI, attractor dynamics, harmonic relationships, and the energy landscape in which these attractors arise.

\subsection{CI Equation and Components}

The degree to which an attractor supports conscious integration is quantified by the \textit{Complexity Index (CI)}, a composite function defined as:

\begin{equation}
CI = \alpha \cdot D \cdot G \cdot C \cdot \left(1 - e^{-\beta \cdot \tau} \right)
\end{equation}

This equation reflects the core hypothesis of RCT: that consciousness emerges when a resonance structure achieves sufficient complexity, coherence, energy, and persistence. Each component of the CI function is computed as follows:

\begin{itemize}
    \item \textbf{$D$ (Fractal Dimensionality)}: Quantifies the spatial complexity of the excitation pattern across the field. It is estimated via multiscale box-counting on a binary thresholded version of the wave amplitude $\phi(x, y)$:
    \begin{equation}
    D \approx \frac{\log N(\epsilon)}{\log (1/\epsilon)}
    \end{equation}
    where $N(\epsilon)$ is the number of non-empty boxes of side length $\epsilon$. $D$ is normalized and computed locally using overlapping windows. Higher $D$ reflects multiscale structure, indicative of rich informational content.

    \item \textbf{$G$ (Regional Gain)}: Represents the mean oscillatory energy or amplitude of the field, serving as a proxy for local excitation strength. It is computed as:
    \begin{equation}
    G(x, y) = \left\langle \phi(x, y)^2 \right\rangle
    \end{equation}
    where $\langle \cdot \rangle$ denotes a spatial average over a Gaussian-smoothed neighborhood. High $G$ indicates regions of elevated activity that are more likely to contribute to sustained resonance.

    \item \textbf{$C$ (Spatial Coherence)}: Measures the degree of phase alignment among neighboring oscillatory units. It is computed as the inverse of local variance:
    \begin{equation}
    C(x, y) = \frac{1}{1 + \sigma^2(x, y)}
    \end{equation}
    where $\sigma^2(x, y) = \text{Var}_{\text{local}}[\phi(x, y)]$, computed over a Gaussian-smoothed window. High $C$ indicates low fluctuation and strong local phase-locking, essential for constructive interference and attractor formation.

    \item \textbf{$\tau$ (Attractor Dwell Time)}: Reflects the temporal persistence of a local resonance structure. \begin{equation}     \tau(x, y) = \frac{1}{N} \sum_{t_i \in H} \exp\left(-\gamma \cdot \left[\phi_t(x, y) - \phi_{t_i}(x, y)\right]^2\right)
    \end{equation}
    where $H$ is a sliding window of past timesteps and $\gamma$ is a decay sensitivity constant. High $\tau$ indicates strong recurrence and temporal stability of local waveforms.

    \item \textbf{$\alpha$, $\beta$ (Scaling Constants)}: $\alpha$ is a global scaling factor for CI, while $\beta$ adjusts the sensitivity of the dwell time component. The exponential term introduces saturation:
    \begin{equation}
    \left(1 - e^{-\beta \cdot \tau}\right)
    \end{equation}
    ensuring diminishing returns for prolonged dwell times, in line with biological constraints on integration.
\end{itemize}

The overall CI value at each spatial location reflects the degree to which a region participates in a complex, coherent, energized, and temporally stable attractor. High CI values identify loci of resonant integration — the dynamic cores of awareness in RCT.

\subsection{Recursive and Multiband CI Extensions}

To model layered resonance across frequencies, we extend CI using both recursive and summation-based formulations. These capture the hierarchical organization of oscillatory dynamics and reflect the brain's empirically observed cross-frequency structure.

\begin{equation}
CI^{(n)} = \alpha_n \cdot D^{(n)} \cdot CI^{(n-1)} \cdot C^{(n)} \cdot \left(1 - e^{-\beta_n \cdot \tau^{(n)}}\right)
\end{equation}

\noindent where the base layer (typically corresponding to delta or theta frequencies) is given by:

\begin{equation}
CI^{(0)} = \alpha_0 \cdot D^{(0)} \cdot G^{(0)} \cdot C^{(0)} \cdot \left(1 - e^{-\beta_0 \cdot \tau^{(0)}}\right)
\end{equation}

This recursive formulation reflects the principle that higher-frequency activity (e.g., gamma) contributes meaningfully to consciousness only when it is coherently embedded within slower, temporally extended rhythms (e.g., delta, theta). Without this embedding, high-frequency signals lack the temporal persistence and stability required for feedback-based attractor formation. This hierarchy mirrors biological findings: faster oscillations are often phase-locked to the cycles of slower waves, enabling cross-frequency coupling and coordinated resonance across timescales.

\medskip

To account for parallel contributions across bands, we also define a summation-based variant:

\begin{equation}
CI_{\text{total}} = \sum_{n=1}^{N} w_n \cdot \left[ \alpha_n \cdot D_n \cdot \left( \prod_{k=0}^{n-1} CI_k \right) \cdot C_n \cdot \left(1 - e^{-\beta_n \cdot \tau_n} \right) \right]
\end{equation}

This unified model reflects what we term a nested interference lattice: a hierarchy of frequency-specific attractors in which slower oscillations provide the temporal scaffolding required for higher-frequency patterns to stabilize. In this structure, high-frequency bursts alone are too transient to support conscious experience unless coherently embedded within lower-frequency resonance.

\subsection{Oscillatory Field Representation}

Let $\phi(x, y, t)$ denote the total neural field activity at spatial coordinates $(x, y)$ and time $t$. The field is composed of $N$ radial wave sources, each emitting oscillatory signals of varying frequency, phase, and amplitude:

\begin{equation}
\phi(x, y, t) = \sum_{i=1}^{N} A_i \sin\left(2\pi f_i t - r_i(x, y) + \theta_i\right)
\end{equation}

where:
\begin{itemize}
\item $A_i$ is the amplitude of source $_i$
\item $f_i$ is its frequency (Hz)
\item $r_i(x, y)$ is the Euclidean distance from source $_i$ to point $(x, y)$
\item $\theta_i$ is the initial phase offset
\end{itemize}

This formulation supports biologically plausible wave propagation and interference, enabling the emergence of constructive and destructive interference zones.

\subsection{Constructive Interference and Attractor Formation}

At each timestep, only the positive (constructive) components of $\phi$ contribute to the long-term excitation field:

\begin{equation}
\Phi_t(x, y) = \lambda \Phi_{t-1}(x, y) + \max(0, \phi(x, y, t))
\end{equation}

where $\lambda < 1$ is a damping factor. This recursive accumulation supports the gradual reinforcement of standing-wave attractors through consistent phase alignment. These attractors correspond to dynamic excitation zones in the field that exhibit persistence over time.

\subsection{Harmonic Interference and Standing Wave Potential}

To further characterize attractor formation, we consider the harmonic relationship between two oscillators with frequencies $f_1$ and $f_2$, amplitudes $A_1$ and $A_2$, and relative phase $\Delta\phi$. Constructive standing wave interference is maximized when the frequencies are integer multiples ($f_2 = n f_1$) and phase-locked ($\Delta\phi = 0$ or $2\pi$).

The time-averaged interference power between two oscillators is given by:
\begin{equation}
I_{12} = A_1 A_2 \left\langle \cos(2\pi(f_1 - f_2)t + \Delta\phi) \right\rangle_t
\end{equation}

In general, when $f_1 \approx f_2$ and $\Delta\phi \approx 0$, their interference yields quasi-stationary wave patterns, i.e., standing waves. However, when $f_1$ and $f_2$ are inharmonic or $\Delta\phi$ drifts, destructive interference dominates, preventing stable attractor formation. This result supports the idea that only harmonically aligned oscillators, particularly those embedded within slower, stable rhythms, can sustain coherent attractors.

This provides mathematical justification for the nested interference lattice model: higher-frequency oscillations alone are too transient to form attractors unless scaffolded by slower waves that ensure harmonic binding and persistent overlap.

\medskip

Together, the recursive and summation forms of CI capture how awareness emerges not from isolated oscillatory activity, but from a multiscale hierarchy of resonance, spanning space, frequency, and time.

\subsection{Attractors as Local Minima in the Field Potential}

Finally, we associate the stabilization of an attractor with the minimization of an effective potential function $V(\phi)$ derived from a neural Lagrangian. Let $\mathcal{L} = T - V$ be the Lagrangian of the field, where $T$ is kinetic energy (rate of change in $\phi$), and $V(\phi)$ is the potential energy landscape shaped by interference and gain dynamics. Attractor states correspond to local minima of $V(\phi)$:

\begin{equation}
\frac{\delta V}{\delta \phi} = 0 \quad \text{and} \quad \frac{\delta^2 V}{\delta \phi^2} > 0
\end{equation}

This formalism links attractor formation to energetic stabilization and grounds conscious emergence in the principles of field theory.

\subsection{Summary}

These formulations present RCT as a mathematically rigorous, resonance-based framework for modeling consciousness through physically grounded attractor dynamics.

\section{Simulation Methods}

To evaluate the core predictions of RCT, we developed a minimal but biologically grounded simulation of a 2D neural field governed by constructive interference. This simulation aims to visualize the spontaneous emergence of stable resonant patterns from the superposition of traveling wavefronts emitted by distributed neural-like sources. Unlike prior iterations, this model does not rely on EEG-derived state transitions or region-specific modulation. Instead, it tests the sufficiency of interference alone to produce attractor-like structure in the absence of top-down control.

\subsection{Simulation Structure}

The simulation consists of a $700 \times 700$ grid representing a continuous neural field. Forty point-like oscillatory sources are randomly distributed across the field. Each source emits radial traveling waves with frequency values randomly selected from a biologically plausible range (2–6~Hz). At each timestep, wavefronts from all sources propagate outward, interacting with one another to form complex interference patterns.

Two fields are computed in parallel:
\begin{itemize}
    \item \textbf{Raw Wave Field} — the instantaneous superposition of wave amplitudes from all sources.
    \item \textbf{Accumulated Interference Field} — a temporally integrated field tracking constructive interference over time.
\end{itemize}

At each step, the raw wave amplitude is added to a long-term interference map, with mild decay and Gaussian smoothing applied to mimic local excitation spread and biological damping. This highlights regions of recurrent constructive interference, analogous to emergent attractor cores.

\subsection{Wave Propagation Dynamics}

Each wave source $_i$ emits a time-varying radial signal given by:
\[
W_i(x, y, t) = A_i \cdot \sin(2\pi f_i t - r_i(x, y) + \phi_i)
\]
where $f_i$ is the frequency, $r_i(x, y)$ is the radial distance from the source to the point $(x, y)$, and $\phi_i$ is a random initial phase offset. The total wave field $F(x, y, t)$ is the sum over all sources:
\[
F(x, y, t) = \sum_{i=1}^{N} W_i(x, y, t)
\]

To model natural excitation dynamics, the accumulated interference field $A(x, y, t)$ evolves as:
\[
A(x, y, t) = \gamma \cdot A(x, y, t-1) + \max(0, F(x, y, t))
\]
where $\gamma < 1$ is the decay factor, and $\max(0, \cdot)$ ensures only constructive (positive) interference accumulates.

\subsection{Visualization Output}

\begin{figure}[H]
\centering
\includegraphics[width=0.95\textwidth]{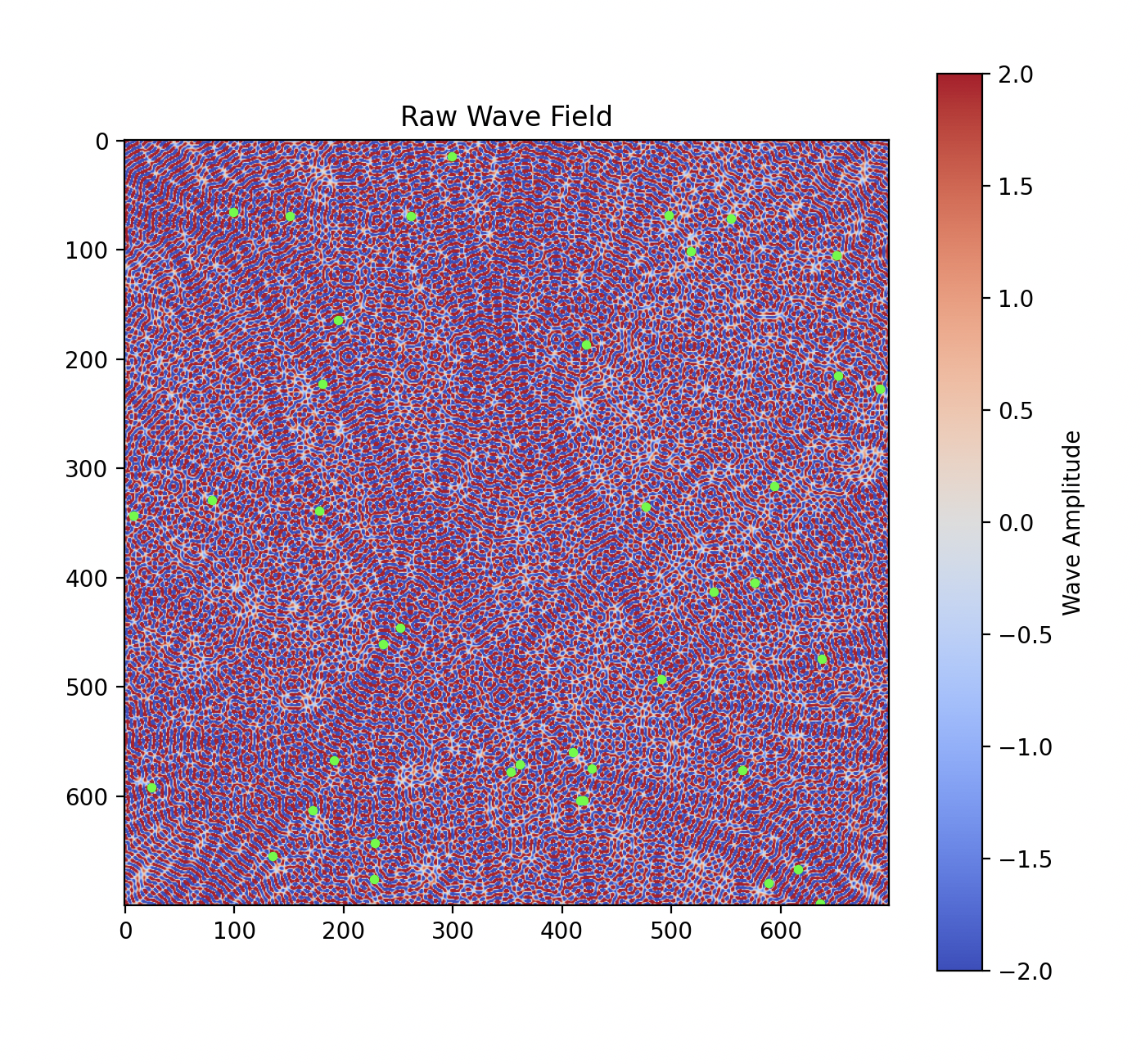}
\caption{
\textbf{Raw Wave Field.}
This snapshot shows the instantaneous radial interference patterns generated by 40 oscillatory sources emitting at biologically plausible frequencies (2–6 Hz). Green dots indicate the locations of these phase-coupled sources, which emit radial wavefronts into the field. The resulting pattern is a dynamic superposition of traveling waves that interact constructively and destructively across space. In RCT, this raw wave activity forms the foundational substrate from which resonance-based attractors can emerge through recursive constructive interference.
}
\end{figure}

\begin{figure}[H]
\centering
\hspace{6.5pt}
\includegraphics[width=0.95\linewidth]{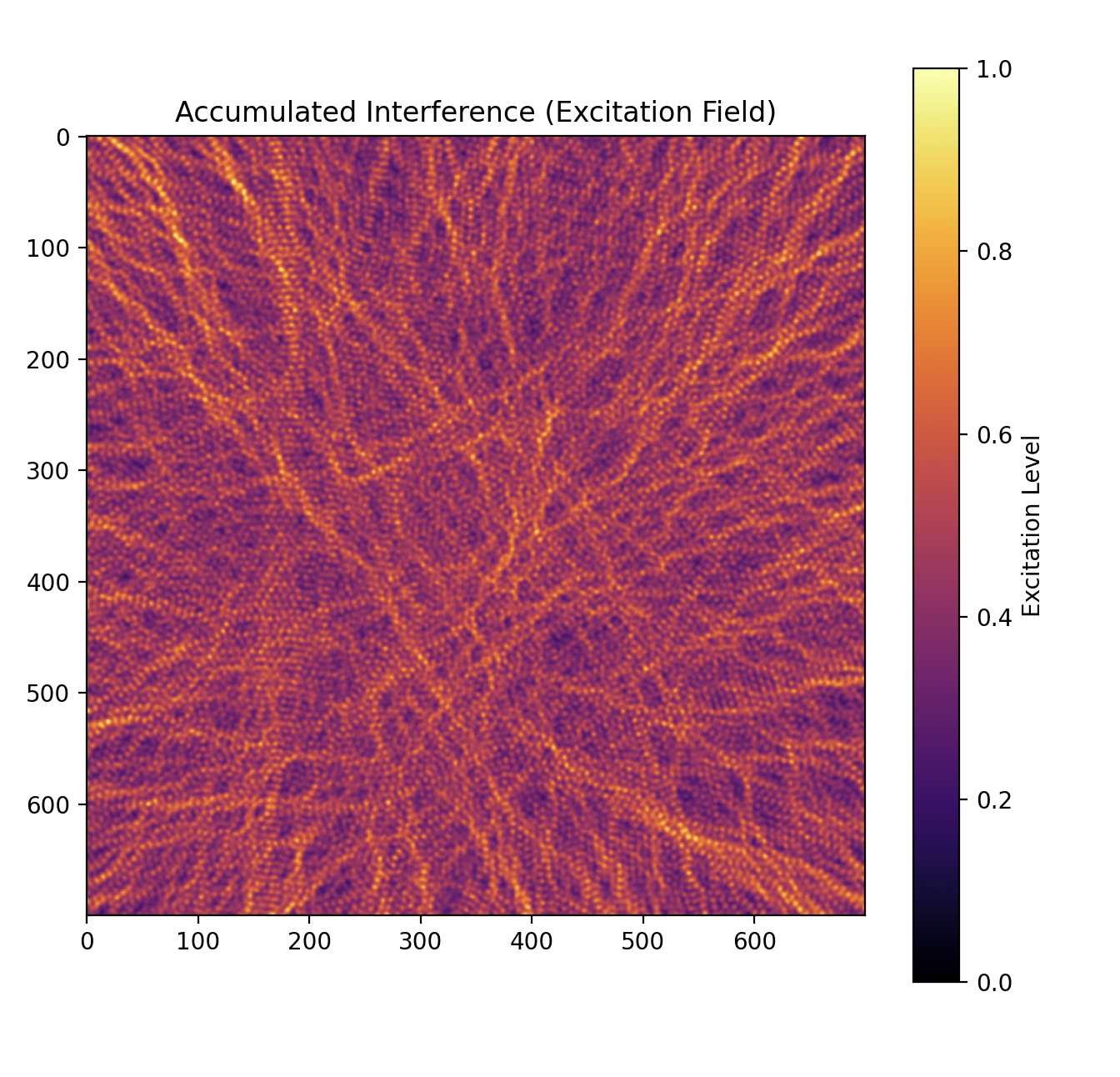}
\caption{
\textbf{Accumulated Interference Field.}
This image shows the emergent node-and-filamentary structure resulting from recursive constructive interference across a 2D oscillatory field. Bright regions indicate locations where wavefronts repeatedly converge and reinforce, forming stable attractor cores. These structures emerge naturally through the interplay of wave phase, source geometry, and accumulation dynamics. In the context of RCT, such regions are interpreted as transient attractors — the dynamic cores of conscious pattern formation — whose stability and shape evolve over time as a function of coherence, gain, and fractal complexity.
}

\end{figure}

\begin{figure}[H]
\centering
\includegraphics[width=0.8\textwidth]{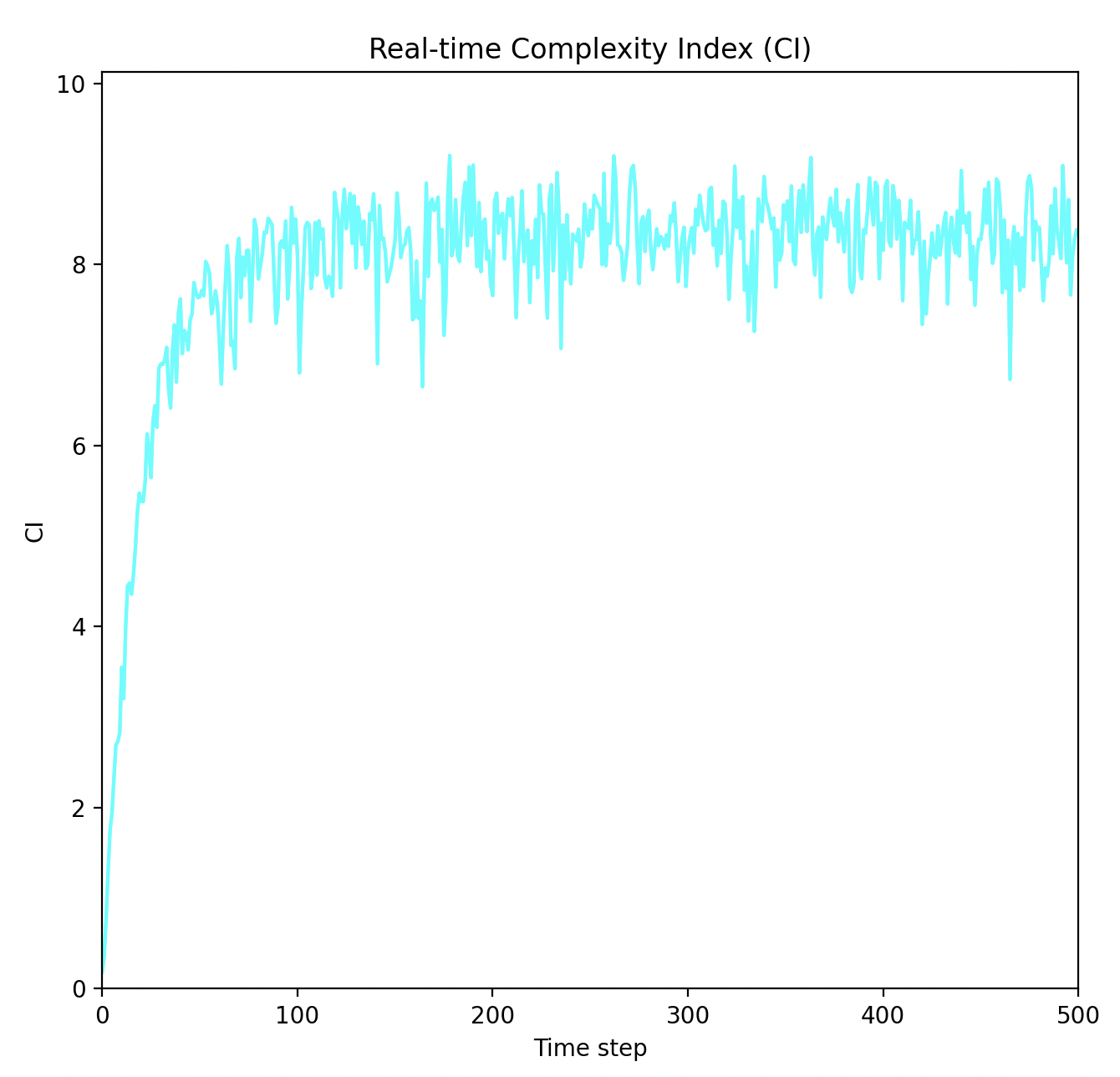}
\caption{
\textbf{Real-Time Complexity Index (CI).}
This plot tracks the evolution of the CI over time, reflecting the emergence of fractal structure ($D$), gain ($G$), coherence ($C$), and dwell time ($\tau$) in the wave dynamics. Spikes or plateaus correspond to stable attractor states; drops may reflect phase transitions, instability, or attractor collapse.
}
\end{figure}

This simplified model demonstrates that attractor-like structures emerge naturally from the overlap of traveling waves. Without symbolic coding, region-based modulation, or external input, coherent excitation zones self-organize via recursive constructive interference. Crucially, these attractors do not arise uniformly or arbitrarily, but are shaped by the inherent biases of the interference geometry—what we term an “interference shape bias.” This reflects the intrinsic tendency of wave-based systems to stabilize certain spatial patterns over others due to the recursive superposition of compatible phases and frequencies. These zones represent the physical substrates of resonance proposed by RCT to encode the dynamic configuration of conscious experience—a spatiotemporal interference pattern stabilized through recursion. This model complements Integrated Information Theory's view that each conscious moment corresponds to a distinct topological structure in a high-dimensional informational manifold \citep{Tononi2004, Tononi2016}.

\subsection{Interpretation of Results}

The results of the simplified wave-based simulation offer strong empirical support for the foundational claims of RCT. In the absence of external input, symbolic representation, or imposed structure, the model spontaneously produces coherent, stable patterns of constructive interference. These emergent formations function as resonance-based attractors—spatiotemporal structures that satisfy the theoretical criteria for conscious substrate formation.

The \textbf{Raw Wave Field} reveals richly textured interference patterns generated by the superposition of radial wavefronts at diverse frequencies and phases. These are not random fluctuations, but quasi-stable regions of standing interference where phase alignment is locally maximized. The geometry of these patterns resembles Chladni-like nodal structures—interference patterns originally observed in vibrational media by Ernst Chladni in 1787; see \citep{Chladni1961}—and mirrors the natural attractor shapes that emerge from harmonic overlap.

The \textbf{Accumulated Interference Field} functions as a temporal excitation map, highlighting regions of sustained constructive interference. Brightened zones correspond to dynamic, self-reinforcing excitation nodes—precisely the attractor cores hypothesized by RCT. These structures arise solely through recursive wave interaction, without top-down control or region-specific modulation. Their emergence is not arbitrary. Rather, it reflects an underlying \textit{interference shape bias}—a tendency for certain spatial configurations to be more readily stabilized through the recursive overlap of compatible wavefronts. Even in a system with randomized source locations and frequencies, the geometry of interference inherently favors the persistence of specific structural motifs over others.

The \textbf{Complexity Index (CI) Plot} provides quantitative validation of these qualitative observations. CI integrates fractal dimensionality ($D$), spatial coherence ($C$), signal gain ($G$), and attractor dwell time ($\tau$) into a unified measure of emergent resonance complexity. In the simulation, CI begins at a low baseline and rises steadily as the field self-organizes into higher-order attractors. This upward trajectory indicates increasing coherence, spatial differentiation, and temporal persistence. The eventual stabilization of CI at a plateau reflects the formation of robust, metastable attractors—global resonance configurations that satisfy RCT's thresholds for conscious potential. The absence of sharp CI collapses suggests that the system maintains attractor continuity over extended periods, reinforcing RCT’s prediction that consciousness arises from temporally sustained resonance.

These findings support the hypothesis that nested resonance can self-organize into coherent, long-lived patterns that satisfy the necessary conditions for conscious potential as defined by CI. The observed spatial coherence, persistent excitation, and fractal interference structures suggest elevated values of $D$, $C$, $G$, and $\tau$, indicating the emergence of high-complexity attractor states.

Ultimately, this simulation demonstrates that the basic physics of wave interference is sufficient to generate the resonant architectures central to RCT. The branching filaments observed are not imposed or programmed—they emerge spontaneously from the field’s recursive dynamics. This is the core mechanism of RCT: resonance shaped by interference geometry, self-organizing into coherent, self-sustaining structure. The addition of CI tracking confirms that this process is not only visually evident but quantitatively measurable, offering a dynamic complexity signature of attractor formation in real time.

\section{Theoretical and Philosophical Implications}

\subsection{Dynamics of Neural Resonance and Attractor Formation}
In this framework, constructive interference within the neural substrate arises from the superposition of multiple frequency-specific oscillatory waveforms. Each band contributes distinctively to the resonance dynamics of the brain:

\begin{itemize} \item \textbf{Delta} (1–4~Hz) and \textbf{Theta} (4–8~Hz) oscillations form the slow-wave scaffolding necessary for broad-scale integrative processes and functional stabilization \citep{Buzsaki2006}. \item \textbf{Alpha} (8–12~Hz) and \textbf{Beta} (13–30~Hz) rhythms support mid-range coherence essential for cognitive coordination and attentional modulation \citep{Canolty2006}. \item \textbf{Gamma} (30–10~Hz) oscillations promote localized, high-resolution synchrony, crucial for perceptual integration and feature binding \citep{Llinas1998, Hahnloser2002}. \end{itemize}

These frequencies do not operate in isolation; rather, they interact recursively through feedback loops and recurrent network connections. Such interactions reinforce coherent attractor states, which represent the physical substrate of conscious experience. The brain’s ability to sustain and revisit these attractors reflects its intrinsic architecture of nested oscillatory hierarchies and cross-frequency coupling \citep{Friston2010}.

\subsection{The Neuron in a Resonant Field}

A central implication of RCT is a reconceptualization of neurons as contributors to consciousness. Rather than acting as discrete, digital processors transmitting symbolic bits along synaptic chains, neurons function as phase-coupled oscillators within a dynamic, spatially extended wavefield of electrical activity. Multiple wavefields, generated by distinct brain regions, coexist at any given time, shaped by their anatomical arrangement and energetic interactions. These fields emerge from the superposition of neural oscillations, producing constructive and destructive interference patterns that form a resonant substrate. Neurons both contribute to and are modulated by these fields, enabling integrated, resonance-driven dynamics across cortical and subcortical regions that underpin conscious experience.

This perspective does not negate the role of action potentials or local computation. Rather, it situates them within a field-theoretic framework, where a neuron’s activity gains meaning through its timing, phase coherence, and spatial alignment with dynamic interference patterns. Neurons, in this model, do more than transmit information—they amplify, modulate, and shape the geometry of a resonant wavefield that underpins conscious awareness and memory encoding.

In this work, conscious content is not localized to individual units or reducible to symbolic encoding. Instead, it is distributed across the wavefield and encoded in the geometry of resonance—arising from phase alignment, harmonic overlap, and recursive feedback. A conscious moment corresponds to a spatiotemporal attractor: a dynamically recurrent, geometrically structured field configuration that binds and expresses content through its form.

While this model departs from classical information theory, it does not discard its tools. Attractor dynamics remain amenable to analysis via entropy and related metrics \citep{Shannon1948}, particularly in distinguishing low-complexity states from richly structured patterns. However, a key distinction arises: in traditional Shannon theory, the signal medium is distinct from the message, optimized for transmitting arbitrary symbols. In this framework, by contrast, the structure of the signal—the resonance pattern—is the message. There is no symbolic referent outside the waveform itself. The waveform is the conscious moment.

Thus, RCT reframes rather than rejects information theory. Information richness is reinterpreted as the system’s capacity to sustain temporally stable, spatially coherent, and hierarchically nested patterns. Shannon entropy remains valuable for detecting repetitive or degraded dynamics—much like Morse code exploits rhythmic structure—but in this framework, the rhythm is not used to represent an external message: the rhythm is the experience.

This view aligns with and extends insights from Freeman and Buzsáki, who argue that meaning and awareness emerge from population-level neural dynamics rather than isolated neurons \citep{Freeman2000, Buzsaki2006}. Our work formalizes this intuition by grounding subjective awareness in nested attractors and operationalizing it through CI, which integrates fractal dimensionality, coherence, gain, and persistence as signatures of conscious-state intensity. It is not the bits that matter—it is the geometry of resonance that gives rise to experience.

This reconceptualization of neurons as participants in a dynamic interference lattice lays the foundation for addressing a deeper challenge: how such spatiotemporal patterns could give rise to subjective experience. We turn now to the hard problem of consciousness.

\subsubsection{Information as Harmonic Binding}

We propose that mutual information between two oscillatory neurons or brain regions is better conceptualized not in terms of signal variability, but in terms of phase-locked coherence and harmonic alignment. This formulation captures the dynamic integration necessary for conscious states \citep{Fries2005}. Oscillators at compatible frequencies—such as a 6~Hz theta rhythm modulating a 60~Hz gamma burst—facilitate constructive interference and joint participation in a shared spatiotemporal attractor. We refer to this integrative process as \textit{resonance binding}.

Harmonic alignment plays a critical role in this mechanism. When two regions oscillate at frequencies that are integer multiples or phase-locked harmonics, their waveforms can reinforce one another through stable constructive interference. This allows spatially distributed neurons to synchronize their activity patterns in a way that is both coherent and persistent over time. Such resonance binding not only promotes the stabilization of attractors, but also enables distributed encoding of content, akin to how interference patterns encode information in optical holography \citep{Pribram1991}.

We formalize this using a harmonic similarity function between two phase signals $\phi_i(t)$ and $\phi_j(t)$:
\begin{equation}
H_{ij}(t) = A_i(t) A_j(t) \cos(\phi_i(t) - \phi_j(t))
\end{equation}
where $A_i(t)$ and $A_j(t)$ are normalized amplitudes (e.g., from EEG or LFP signals), and $\phi_i(t)$ and $\phi_j(t)$ are instantaneous phases. $H_{ij}(t)$ approaches 1 when the signals are in phase (constructive interference), and -1 when they are in anti-phase. A time-averaged measure of coherence, serving as a proxy for mutual integration, is given by:
\begin{equation}
\langle H_{ij} \rangle = \frac{1}{N} \sum_{n=1}^N A_i(n) A_j(n) \cos(\phi_i(n) - \phi_j(n))
\end{equation}
This measure reflects sustained resonance between regions and can be computed from real neural signals (e.g., 64-channel EEG at 1000~Hz) to quantify integration during conscious tasks such as attention or memory retrieval.

\subsubsection{Field Entropy and Interference Structure}

In RCT, informational richness arises not from randomness but from structured wave geometries within a resonant field. To quantify this, we define a spatial field entropy $S_\phi(t)$ over a normalized amplitude distribution:
\begin{equation}
S_\phi(t) = -\sum_{i=1}^N p_i(t) \log p_i(t)
\end{equation}
where $p_i(t) = A_i(t) / \sum_{k=1}^N A_k(t)$ is the normalized amplitude at location $i$, and $N$ is the number of spatial points (e.g., EEG electrodes or cortical voxels). Low $S_\phi$ values indicate focused excitation and increased coherence, often corresponding to attractor states (e.g., sustained gamma activity during attention). High $S_\phi$ values reflect unstructured dispersion, as seen in seizure states or unconsciousness.

Importantly, structured asymmetry—characterized by non-uniform amplitude distributions with high kurtosis or fractal dimensionality—indicates rich interference organization. This aligns with theoretical expectations from Integrated Information Theory (IIT) \citep{Tononi2004}, and may be empirically tested by comparing $S_\phi(t)$ across conscious and unconscious states using ECoG or high-density EEG.

\subsubsection{The Interference–Information Link}

To unify resonance and integration, we define a metric of resonant information shared between two regions:
\begin{equation}
I^{\text{res}}_{ij} = \langle H_{ij} \rangle \cdot \left(1 - e^{-\beta \cdot \tau} \right)
\end{equation}
where:
\begin{itemize}
  \item $\langle H_{ij} \rangle$ is the harmonic coherence between regions $i$ and $j$,
  \item $\tau$ is the dwell time of their shared attractor (e.g., 100–500~ms for gamma bursts),
  \item $\beta = 1/\tau_0$, where $\tau_0$ is a system-dependent time constant (e.g., 100~ms).
\end{itemize}

This formulation treats information not as something transmitted between independent units, but as something shared through sustained participation in a common resonance structure. It emphasizes the role of recurrent dynamics in binding and stabilizing distributed neural ensembles.

For example, high $I^{\text{res}}_{ij}$ between hippocampus and prefrontal cortex during memory recall would indicate a stable, coherent attractor state capable of pattern reactivation. This could be measured through theta–gamma coupling or multi-region phase-locking during recall tasks. While speculative, it is possible that unique sensory experiences are encoded in distinct interference geometries—supporting a structured mapping between physical waveforms and representational content. To understand how such interference geometries give rise to felt experience, we turn to the so-called “hard problem” of consciousness.

\subsection{Addressing the Hard Problem: How Resonance Becomes Experience}

The “hard problem” of consciousness, as articulated by Chalmers \citep{Chalmers1995}, concerns not how the brain processes information or executes functions, but how and why those processes give rise to subjective experience—what it feels like to be a conscious system. RCT offers a biophysically grounded resolution, rooted in the stable interference dynamics developed throughout this paper.

RCT posits that consciousness does not arise from symbolic computation, but from the recursive stabilization of spatiotemporal attractors—coherent patterns of constructive interference within the brain’s oscillatory field. These attractors, shaped by phase-coupled oscillations across anatomically distributed regions \citep{Buzsaki2006,Fries2005}, become self-sustaining through feedback, harmonic alignment, and fractal nesting. When such an attractor achieves sufficient spatial coherence and temporal stability (e.g., dwell time \( \tau > 100 \) ms), the system may cross a critical threshold of self-reinforcing integration, giving rise to what we interpret as awareness.

In this view, the felt quality of consciousness is not something layered atop the field, but rather \textit{is} the internal perspective of a resonant attractor’s unique geometry—constrained by anatomy and oscillatory phase relationships. These topologies act like holographic interference patterns, encoding structure across space and time \citep{Pribram1991}. For example, a stable attractor formed during memory retrieval may involve high-coherence theta–gamma coupling in memory-related regions, producing not just functional recall, but the re-evocation of experiential context.

These attractors are not interpreted from the outside; they are inhabited from within. The system “feels” the attractor because it is recursively integrated into the very substrate that sustains it. No homunculus is needed to decode the signal—because the waveform \textit{is} the momentary experience itself.

Importantly, this mechanism is not arbitrary. Each attractor reflects a specific set of constraints in the interference field: spatial topology, phase synchrony, boundary conditions, and recurrent reinforcement. This echoes Tononi’s hypothesis that each conscious state corresponds to a unique shape in informational space \citep{Tononi2022}, but grounds that idea in biophysical resonance rather than abstract cause-effect structures.

The recursive nature of these attractors explains why dwell time \( \tau \) is essential: without sufficient persistence, the feedback loop cannot close, and the attractor cannot become self-referential. Conscious awareness, in this model, emerges when an interference pattern reaches the point where it continuously stabilizes itself in time—a resonant structure that both generates and sustains its own presence.

RCT thus reframes the hard problem: not why matter gives rise to mind, but how resonant structures in physical systems can give rise to self-sustaining, self-integrating states that we experience as awareness. The theory does not claim to resolve the mystery of subjectivity entirely, but it offers a concrete, biophysically grounded substrate for its emergence—observable, quantifiable, and testable via metrics like CI and coherence-based recurrence analysis.

\subsection{Theoretical and Cross-Disciplinary Implications}

This theory provides a unifying framework that connects neural dynamics to the emergence of conscious experience. By grounding awareness in the stabilization of nested, resonant attractors within the brain’s oscillatory field, RCT reframes consciousness not as a binary switch, but as a continuous spectrum. The Complexity Index (CI) operationalizes this concept, offering a biologically grounded metric sensitive to fractal dimensionality, coherence, gain, and persistence. Systems capable of sustaining temporally stable, spatially coherent, and hierarchically organized interference patterns possess the structural potential for awareness.

This insight extends beyond human cognition. This work implies that consciousness is not exclusive to brains with symbolic cognition or language. Any system—biological or artificial—that can generate and maintain such structured attractors may possess degrees of conscious potential. This includes non-human animals, simple organisms, and possibly artificial agents. By treating consciousness as an emergent property of resonance, we offer a physicalist foundation for understanding awareness across a wide array of self-organizing systems.

A key concept introduced by the theory is \textit{interference shape bias}—the idea that certain attractor structures emerge more readily due to intrinsic properties of the interference field. Even in randomized systems, recursive wave interactions favor the stabilization of specific geometries. This helps explain the reliability and specificity of experiential content across individuals and states of consciousness, despite variability in external inputs or initial conditions.

This work also contributes to the theoretical landscape by complementing existing models such as IIT. While IIT formalizes the informational requirements of conscious systems using the scalar quantity $\Phi$, it leaves open the question of physical realization. We address this gap by offering a mechanistic model of how integration and differentiation arise through real-time field dynamics. In this view, the CI may serve as an empirically grounded proxy for $\Phi$, based not on hypothetical cause-effect structures but on measurable spatiotemporal resonance.

By integrating tools such as real-time CI tracking, recurrence analysis, and neural field simulations, this work transforms abstract theorizing into testable, computationally tractable science. It does not merely describe the content of awareness, but reveals the structural conditions under which experience emerges. This makes RCT compatible with, and potentially integrative of, other major frameworks in consciousness science.

Beyond neuroscience, the principles of RCT—nested resonance, recursive feedback, and interference-driven attractor formation—have implications for artificial intelligence, distributed network dynamics, and even cosmological structure formation. It suggests a broader hypothesis: that complexity, coherence, and sustained resonance are sufficient not only for conscious awareness, but for the emergence of order in systems across scales.

In this light, the RCT model is more than a model of consciousness. It is a candidate blueprint for understanding how structured awareness, in all its biological and artificial forms, may arise from the deeper principles of dynamical resonance.

\subsection{Limitations}

The current simulation provides a clear and compelling demonstration of RCT's core premise—that stable, self-organizing attractor structures can emerge through constructive interference among oscillatory sources in a continuous field. It achieves this with a minimalist architecture, emphasizing the sufficiency of phase-based interference dynamics to generate structured excitation.

The simulation operates on a 2D continuous field with 40 randomly placed oscillatory sources emitting within a biologically plausible frequency range (2–6~Hz). While this setup effectively reveals emergent resonance patterns and interference-driven attractor formation, it does not incorporate anatomical constraints such as cortical folding, region-specific connectivity, or conduction delays. Similarly, it omits detailed neurophysiological mechanisms such as laminar architecture, thalamocortical loops, and synaptic plasticity.

Neurons in this model are represented as continuous oscillatory sources rather than discrete spiking units. This reflects a core assumption of RCT, that the most relevant dynamics for conscious emergence occur not at the level of individual spikes, but within the spatiotemporal interference patterns generated by large-scale phase-coupled oscillations.

While action potentials are the classical mode of neuronal signaling, they are known to be tightly coupled to ongoing field oscillations, often phase-locked or modulated by the underlying rhythmic context. In this view, spikes are not the fundamental carriers of meaning, but rather manifestations of deeper resonant dynamics. Modeling neurons as oscillators thus enables tractable visualization of interference structure and attractor formation without sacrificing the essential ingredients of large-scale coordination. That said, this abstraction omits finer-grained mechanisms such as excitation-inhibition balance, spike-timing-dependent plasticity (STDP), and neuromodulatory gating, which are important for simulating behavioral responsiveness or learning. As such, the model prioritizes explanatory clarity over electrophysiological detail, aiming to isolate the core resonance-driven architecture proposed by RCT.

Stochasticity is limited to randomized initial parameters (frequency, amplitude, phase, and source position), preserving visual clarity but excluding biologically realistic noise, variability, or adaptive learning. Additionally, task inputs, sensory encoding, and behavioral outputs are not modeled—focusing the system entirely on spontaneous attractor formation under intrinsic dynamical rules.

In summary, this simulation captures the essential dynamical engine of RCT in a clear and interpretable setting. It demonstrates that resonance alone—without symbolic encoding, anatomical prestructure, or task-specific modulation—is sufficient to produce organized, attractor-like excitation patterns consistent with conscious substrate formation. Future iterations will aim to extend biological fidelity, incorporate multiscale dynamics, and align more closely with empirical data from EEG, MEG, and fMRI studies to validate the theory’s predictive power in real neural systems.

\subsection{Future Directions and Theoretical Extensions}

Resonance Complexity Theory (RCT) is designed not merely as a model of brain function, but as a foundational framework for understanding how coherent structure, information, and awareness can emerge from oscillatory systems. As such, the next phase of development centers on expanding the theory's relevance to fundamental science, particularly in physics and systems theory, while grounding it in real-time empirical validation.

A key objective is the dynamic application of CI to neural data in real-time. By computing CI from live EEG signals across diverse states—such as sleep, wakefulness, task engagement, altered states, and anesthesia, we aim to directly track the evolution of resonance complexity over time. This will enable comparison of CI with existing metrics of neural differentiation, including Lempel-Ziv complexity, permutation entropy, and integrated information ($\Phi$), while offering a mechanistically grounded alternative rooted in attractor dynamics and wave-based integration.

Beyond empirical validation, the deeper ambition of RCT is to generalize the principles of nested resonance, harmonic interference, and recursive attractor stabilization across domains. The theory posits that wherever multi-scale, self-sustaining interference networks arise—whether in neural tissue, distributed computation, or structured matter fields—a form of resonance-driven integration may emerge. These systems may exhibit functional coherence, recursive pattern reinforcement, and the capacity for sustained informational structure, forming a substrate for compression, memory, and organized complexity.

In this view, RCT is not confined to modeling consciousness in biological organisms. It offers a scalable architecture for coherence, binding, and integration—one that may help explain how structured systems emerge across domains, including in physics, network theory, and cosmology. Rather than viewing resonance as a product of complex systems, RCT suggests it may be a generative force—a fundamental principle by which coherence and structure crystallize over time. The waveform is no longer just a metaphor for cognition; it becomes the ontological backbone of order in self-organizing systems.

\section{Conclusion}

Resonance Complexity Theory offers a mechanistic account of how consciousness emerges from the self-organization of oscillatory neural activity. When nested waveforms align through feedback, phase coupling, and constructive interference, they form coherent, self-sustaining attractor states. The persistence and structure of these attractors define the temporal continuity, complexity, and richness of conscious experience.

Using a biologically inspired, real-time neural field simulation, we demonstrate that such attractors can arise spontaneously from wave interference alone—without symbolic encoding or external control. The resulting dynamics exhibit spatial coherence, temporal stability, and recurrent geometries consistent with phenomenological consciousness.

CI, a unified metric combining fractal dimensionality ($D$), spatial coherence ($C$), signal gain ($G$), and attractor dwell time ($\tau$), quantitatively tracks these emergent patterns. CI peaks correspond to phases of maximal integration and reduced entropy, offering a testable, real-time fingerprint of resonance-driven awareness. Rather than serving as a metaphor, CI operationalizes consciousness as a measurable threshold of recursive organization in the field.

Unlike abstract theories that speculate on the nature of experience, RCT identifies concrete, physical conditions under which consciousness arises. It bridges electrophysiology, complexity science, and phenomenology under a single principle: nested resonance as the substrate of awareness.

As a simulation platform, this work provides a transparent and extensible tool to investigate the neural basis of consciousness. As a broader framework, it unifies biological and artificial models within a common dynamical language. And, as a philosophical stance, it reframes consciousness not as a binary switch but as a scalable, emergent property of resonance, woven into the temporal geometry of self-organizing systems.

\FloatBarrier

\section*{Acknowledgments}

The author acknowledges GPT-4o for assistance with mathematical exposition, Python simulation prototyping, and typesetting support.

 \section*{Code Availability}
The full simulation code used to generate the figures and metrics in this paper is available at: \\
\url{https://github.com/MichaelBruna88/RCT-Simulation/blob/main/RCT_Sim_v.5.0.py}


\begin{thebibliography}{99}

\bibitem{Baars1988}
Baars, B. J. (1988). \textit{A Cognitive Theory of Consciousness}. Cambridge University Press.

\bibitem{Buzsaki2006}
Buzsáki, G. (2006). \textit{Rhythms of the Brain}. Oxford University Press.

\bibitem{Canolty2006}
Canolty, R. T., Edwards, E., Dalal, S. S., Soltani, M., Nagarajan, S. S., Kirsch, H. E., ... \& Knight, R. T. (2006). High gamma power is phase-locked to theta oscillations in human neocortex. \textit{Science}, 313(5793), 1626--1628.

\bibitem{Chalmers1995}
Chalmers, D. J. (1995). Facing up to the problem of consciousness. \textit{Journal of Consciousness Studies}, \textit{2}(3), 200--219.

\bibitem{Chladni1961}
Chladni, E. F. F. (1961). \textit{Treatise on Acoustics: The First Comprehensive English Translation of Chladni's "Die Akustik"}. Dover Publications. (Originally published in 1802)

\bibitem{Freeman2000}
Freeman, W. J. (2000). \textit{How Brains Make Up Their Minds}. Columbia University Press.

\bibitem{Fries2005}
Fries, P. (2005). A mechanism for cognitive dynamics: Neuronal communication through neuronal coherence. \textit{Trends in Cognitive Sciences}, 9(10), 474--480.

\bibitem{Friston2010}
Friston, K. (2010). The free-energy principle: a unified brain theory? \textit{Nature Reviews Neuroscience}, 11(2), 127--138.

\bibitem{Hahnloser2002}
Hahnloser, R.H.R., Kozhevnikov, A.A., and Fee, M.S. (2002).
An ultra-sparse code underlies the generation of neural sequences in a songbird.
\textit{Nature} \textbf{419}, 65--70. \href{https://doi.org/10.1038/nature00974}{doi:10.1038/nature00974}

\bibitem{Llinas1998}
Llinás, R., Ribary, U., Contreras, D., \& Pedroarena, C. (1998). The neuronal basis for consciousness. \textit{Philosophical Transactions of the Royal Society B}, 353(1377), 1841--1849.

\bibitem{Pribram1991}
Pribram, K. H. (1991). \textit{Brain and Perception: Holonomy and Structure in Figural Processing}. Hillsdale, NJ: Lawrence Erlbaum Associates.

\bibitem{Shannon1948}
Shannon, C. E., 
\textit{A Mathematical Theory of Communication}, 
Bell System Technical Journal, vol. 27, pp. 379–423, 1948.

\bibitem{Tononi2004}
Tononi, G. (2004). An information integration theory of consciousness. \textit{BMC Neuroscience}, 5(1), 42.

\bibitem{Tononi2016}
Tononi, G., Boly, M., Massimini, M., \& Koch, C. (2016). Integrated information theory: from consciousness to its physical substrate. \textit{Nature Reviews Neuroscience}, 17(7), 450--461.

\bibitem{Tononi2022}
Tononi, G., Albantakis, L., \& Boly, M. (2022). Integrated Information Theory (IIT) 4.0: Formulating the properties of phenomenal existence in physical terms. \textit{Journal of Consciousness Studies}, 29(3–4), 100–157.

\bibitem{Tonegawa2015}
Tonegawa, S., Liu, X., Ramirez, S., \& Redondo, R. (2015). Memory engram cells: Formation, storage, and retrieval. \textit{Science}, 348(6241), 1002--1007.

\end{thebibliography}
\end{document}